\documentclass[table]{ccjnl}
\usepackage{lipsum,amsmath}
\usepackage{cuted}
\usepackage{bm}
\usepackage{booktabs}
\usepackage{multirow}
\usepackage{makecell}
\usepackage{threeparttable}
\usepackage{booktabs}
\usepackage{array}
\graphicspath{{figures/}}
\title{AI Enlightens Wireless Communication: Analyses, Solutions and Opportunities on CSI Feedback}
\author{Han Xiao\inst{1}, Zhiqin Wang\inst{2,*}, Wenqiang Tian\inst{1}, Xiaofeng Liu\inst{2}, Wendong Liu\inst{1}, Shi Jin\inst{3}, Jia Shen\inst{1}, Zhi Zhang\inst{1}, Ning Yang\inst{1}
}
\receiveddate{}
\reviseddate{}
\Editor{}

\address[1]{Dept.of Standards Research, OPPO, Beijing, China}
\address[2]{China Academy of Information and Communications Technology, Beijing, China}
\address[3]{National Mobile Communications Research Laboratory, Southeast University, Nanjing, China}
\address[*]{Corresponding Author, E-mail:zhiqin.wang@caict.ac.cn}

\begin{document}

\maketitle

\begin{abstract}
In this paper, we give a systematic description of the 1st Wireless Communication Artificial Intelligence (AI) Competition (WAIC) which is hosted by IMT-2020(5G) Promotion Group 5G+AI Work Group. Firstly, the framework of full channel state information (F-CSI) feedback problem and its corresponding channel dataset are provided. Then the enhancing schemes for DL-based F-CSI feedback including i) channel data analysis and preprocessing, ii) neural network design and iii) quantization enhancement are elaborated. The final competition results composed of different enhancing schemes are presented. Based on the valuable experience of 1st WAIC, we also list some challenges and potential study areas for the design of AI-based wireless communication systems. 
\end{abstract}

\section{Introduction}
\label{sec_introduction}
For a wireless communication system, physical layer is the basis of ensuring the quality of communication services. Recently, massive multiple-input multiple-output (MIMO) is considered to be a key physical layer technology for the fifth generation
(5G) communication system to meet the growing needs of the mobile data traffic. Specifically, it is quite essential for MIMO to make effective channel state information (CSI) feedback so that the base station (BS) can accurately determine the channel quality and perform further resource allocation and data transmission.

To achieve channel information extraction and CSI feedback, Type 1 and Type 2 codebooks and corresponding CSI feedback mechanisms have been defined  and standardized in 5G new radio (NR) specification by 3rd Generation Partnership Project (3GPP) \cite{1,2,3}. It is noted that the current effective CSI feedback mechanism reports a lossy channel information, since it performs CSI extraction from the full channel state information (F-CSI) considering the difficulty of compressing and reporting F-CSI using tradition methods as well as the implementation and commercial deployment. Therefore, how to perform F-CSI feedback remains a challenging and attractive issue.

In recent years, with the evolution and improvement of Artificial Intelligence (AI) technology, the nonlinear compression and fitting scheme based on deep learning (DL) has achieved great results both in academia and industry \cite{wang2017deep,wen2018deep,sun2020ancinet,lu2020multi,chen2020deep,mashhadi2020distributed,cao2021lightweight,guo2020deep,guo2021canet,lu2018mimo,wang2018deep,guo2020convolutional,li2020spatio,chen2019novel,lu2019bit,jiang2017end}, which makes the using of full channel state information (F-CSI) becoming possible and could be a potential candidate technical solution for mobile communication system in the future. Among these works, a DL-based F-CSI feedback scheme with convolutional neural network (CNN) model is firstly proposed \cite{wen2018deep}, referred to as the CsiNet, which introduces image compression and recovery techniques to massive MIMO by performing transformation from F-CSI to a codeword at the encoder and inverse transformation at the decoder. After that, a series of follow-up studies have been carried out \cite{sun2020ancinet,lu2020multi,chen2020deep,mashhadi2020distributed,cao2021lightweight,guo2020deep,guo2021canet}, which design various kinds of CNNs to handle different problems in F-CSI feedback, such as an AnciNet to enhance the F-CSI feedback and recovery performance with denoising module \cite{sun2020ancinet}, a DeepCMC composed of convolutional layers followed by quantization and entropy coding blocks \cite{mashhadi2020distributed}, a ConvCsiNet along with ShuffleCsiNet to improve the F-CSI reconstruction performance with limited memory space and computing power \cite{cao2021lightweight}, a CsiFBnet to maximize the beamforming gain at the BS instead of the CSI feedback accuracy \cite{guo2020deep}, and a CANet with uplink-aided downlink F-CSI acquisition mechanism \cite{guo2021canet}. Furthermore, considering the temporal correlations of channels, CsiNet combined with long-short time memory (LSTM) architectures are proposed in \cite{lu2018mimo,wang2018deep,guo2020convolutional,li2020spatio,chen2019novel,lu2019bit,jiang2017end} to take full advantage of the previously channel information. In addition, discussions on DL enhanced quantization as well as dequantization have been conducted in \cite{chen2020deep,chen2019novel,lu2019bit} to improve their performance with limited feedback bits. 
%Moreover, the National Artificial Intelligence Competition (NAIC) is hold by Peng Cheng Laboratory to evaluate the DL-based F-CSI feedback performance with practical channel measurement \cite{guo2020ai}. 
Researches show that DL based approaches provide an attractive and potential way to enhance the performance of F-CSI feedback. Compared with the traditional methods, the actual channel characteristics can be well compressed at the user equiment (UE) side and recovered at the base station (BS) with relatively lower overhead with the help of non-linear feature extraction and fitting capability provided by DL.
\raggedbottom
%In order to further explore the practical application of AI in communication systems, especially to evaluate the performance of DL-based F-CSI feedback, IMT-2020(5G) Promotion Group 5G+AI Work Group held the 1st Wireless Communication AI Competition (WAIC) in spring of 2021 with a topic of Performance Improvement of Deep Learning Based F-CSI Feedback, which aims to evaluate the performance of DL based F-CSI recovery and feedback overhead reduction with data-set on 3GPP system-level channel model. On the one hand, the competition can gather forces from all sides to implement AI-based solutions on conventional communication problems. On the other hand, it can also explore the performance boundaries of AI-based solutions under classical data sets.

The structure of this paper as following. The 1st Wireless Communication AI Competition (WAIC) with the task of DL-based CSI feedback is introduced in Section \ref{WAIC} and the system model and channel model along with the dataset illustration involved in the 1st WAIC are depicted in Section \ref{System Model and Channel Model}. The data preprocessing, model design, quantization enhancement and other related key points are discussed in Section \ref{Enhancing Schemes}. Competition results with some analyses and challenges for DL-based wireless communication are given in Section \ref{Capability Composition of Participating Models}. Finally, a brief conclusion is provided in Section \ref{CONCLUSION}.

%The structure of this paper as following. The system model and channel model along with the dataset illustration involved in the 1st WAIC are depicted in Section \ref{System Model and Channel Model}. The data preprocessing, model design, quantization enhancement and other related key points are discussed in Section \ref{Enhancing Schemes}. Competition results with some analyses and challenges for DL-based wireless communication are given in Section \ref{Capability Composition of Participating Models}. Finally, a brief conclusion is provided in Section \ref{CONCLUSION}.
\vspace{-10.8pt}
\section{Wireless Communication AI Competition}
\label{WAIC}
In order to further explore the practical application of AI in communication systems, especially to evaluate the performance of DL-based F-CSI feedback, IMT-2020(5G) Promotion Group 5G+AI Work Group held the 1st WAIC in the spring of 2021 with a topic of AI Enlightens Wireless Communication which is committed to promoting the deep integration and mutual promotion of the wireless communication and AI. The 1st WAIC focus on the task of Performance Improvement of Deep Learning Based F-CSI Feedback, which aims to evaluate the performance of DL based F-CSI recovery and feedback overhead reduction with classic  data-set on 3GPP system-level channel model. 

There are more than $900$ teams involving $1175$ contestants from related $210$ companies, $160$ universities and research institutes participate in the competition. Students from worldwide universities are full of enthusiasm and account for $61.62\%$ of the participants. Since the high degree of integration between the task and the industry, the remaining $38.38\%$ participants from companies are more than the ones compared to other similar competitions. Focusing on the schedule of the competition, the two-month online results submission for participants and the one-day offline seminar for intrested researchers provide a good opportunity for technical exploring and sharing.

In this artical, we provide the details of dataset construction method and the noteworthy enhancing schemes for F-CSI feedback in the 1st WAIC. The results with a component analysis are also listed. On the one hand, the competition gather forces from all sides to implement AI-based solutions on conventional communication problems. On the other hand, it also explore the performance boundaries of AI-based solutions under classical data sets.

\vspace{8pt}
\section{System Model and Channel Model}
\label{System Model and Channel Model}
\subsection{System Model}

\begin{figure}[!tbp]
\centering
\includegraphics[width=1\columnwidth]{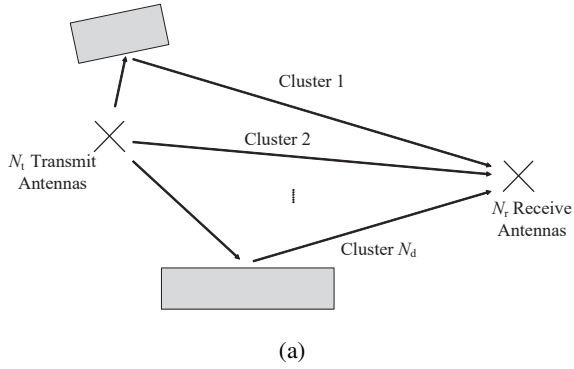} \\
\text{\footnotesize (a)}
\includegraphics[width=1\columnwidth]{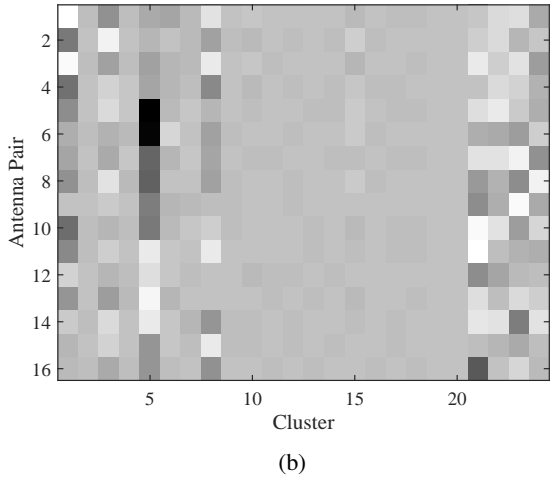}
\text{\footnotesize (b)}
 \caption{Illustration of (a) multi-path channel model and (b) gray scale of energy for one channel sample.}
\label{fig1}
\end{figure}

As shown in Fig. \ref{fig1} (a), a typical MIMO system with $N_{\rm t}$ transmit antennas at the BS and $N_{\rm r}$ receive antennas is considered. A channel $\mathbf{H}$ with $N_{\rm d}$ clusters needs to be compressed on the UE side and recovered at the BS, which can be denoted as
\begin{equation}
\mathbf{H}=\left[\mathbf{H}_1,\mathbf{H}_2,\cdots,\mathbf{H}_{N_{\rm d}}\right]
\end{equation}
where
\begin{equation}
\mathbf{H}_d = \left[\begin{array}{lll}
h_{1,1,d} & \cdots & h_{1,N_{\rm r},d} \\
\vdots & \ddots & \vdots \\
h_{N_{\rm t},1,d} & \cdots & h_{N_{\rm t},N_{\rm r},d}
\end{array}\right]
\vspace{5pt}
\end{equation}
and $h_{i,j,d}$ denotes the channel coefficient of the $i$-th transmit antenna, $j$-th receive antenna and $d$-th cluster. Consequently, the total number of feedback coefficients turns out to be $N=N_{\rm d}\times N_{\rm t}\times N_{\rm r}$. The gray scale of $\mathbf{H}$ is also described in Fig. \ref{fig1} (b), where x-axis indicates the cluster index and y-axis indicates the antenna pair index, and deeper color means the larger energy of the corresponding channel element.

\begin{figure}[tbp]
 \begin{center}
 \includegraphics[width=1\columnwidth]{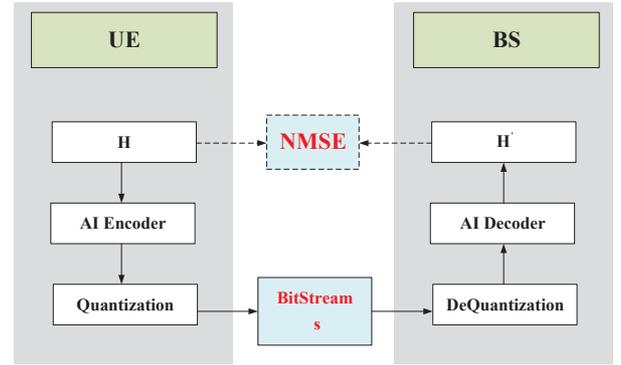}
 \end{center}
% \vspace*{-5mm}
 \caption{Illustration of DL-based F-CSI feedback mechanism.}
 \label{fig2}
% \vspace*{-5mm}
\end{figure}

A DL-based F-CSI feedback mechanism is shown in Fig. \ref{fig2}, where a UE encodes $\mathbf{H}$ to a $M$-dimensional bitstream $\mathbf{s}$ through an AI-Encoder and forwards it to the BS, and the BS decodes the received bitstream $\mathbf{s}$ to $\mathbf{H}'$ through an AI-decoder to recover the F-CSI. After that, the codebook matching and UE scheduling can be performed accordingly based on $\mathbf{H}'$ at the BS. A normalized MSE is used as the criterion to evaluate the difference between the recovered channel $\mathbf{H}'$ and the original channel $\mathbf{H}$, which is quantified by
\begin{equation}
\vspace{5pt}
\rm{NMSE} = \rm{E}\left\{\frac{||\mathbf{H}'-\mathbf{H}||_2^2}{||\mathbf{H}||_2^2}\right\}
\vspace{3pt}
\end{equation}
Most of the current researches on F-CSI feedback focus on the performance of NMSE. However, in the actual system design, it is more important to reduce the feedback overhead as much as possible with a fixed NMSE level. Therefore, the goal of the 1st WAIC competition is to minimize the feedback overhead with a NMSE threshold constraint. Considering the size of dataset and the requirement of channel feedback accuracy in actual system, the threshold of NMSE is set as $0.1$. Namely, only the schemes with NMSE $\leq 0.1$ can be treated as an effective solution.

\vspace{5pt}
\subsection{Channel Model}
\vspace{5pt}
In order to reflect the multiple and complex channel characteristics in realistic environments to some extent, the system-level three-dimensional (3D) channel model is considered, which has been a benchmark channel for simulation and performance evaluation in 3GPP. Generally, based on the various deployment scenarios of massive MIMO, four kinds of channel models, including urban-macro (UMa), urban-micro (UMi), indoor hotspot (InH) and rural macro (RMa) have been abstracted and defined by 3GPP \cite{4}. Specifically, the UMa scenario is utilized in our work. Moreover, the pure non-line-of-sight (NLoS) channels without line-of-sight (LoS) components are utilized, wherein the system-level downlink channel matrix $\mathbf{H}_{\rm sys}$ can be given as
\begin{equation}
\mathbf{H}_{\rm sys} = \sum_{d=1}^{N_{\rm d}} \mathbf{H}_d = \sum_{d=1}^{N_{\rm d}}\sum_{l=1}^{L_d} \mathbf{H}_{d,l}
\label{eqHsys}
\end{equation}
where $\mathbf{H}_{d,l}$ and $L_d$ indicate the $l$-th sub-path channel and the total number of sub-paths in the $d$-th cluster, respectively. Ignoring the large-scale influence, such as path-loss and outdoor-to-indoor loss, only small-scale fading is considered, where each sub-path $\mathbf{H}_{d,l}$ is generated based on \cite{4} with different angle-of-departures/arrivals, angle-spreads in both azimuth and zenith domains, power and delay distributions, and initial phases. More specific configurations about the channel models can be found in \cite{4}.

\vspace{8pt}
As for the dataset, $N_{\rm slot}\times N_{\rm UE}$ channel samples are provided, wherein $N_{\rm slot}$ slots are uniformly sampled with $T$ continuous interval slots for each UE, $N_{\rm UE} = N_{\rm train} + N_{\rm test}$ is the number of randomly distributed UEs in $N_{\rm c}$ cells. Specifically, $N_{\rm train}$ and $N_{\rm test}$ represent the numbers of UEs in training set and test set, respectively. The corresponding channel and dataset parameter settings are listd in Table \ref{tabChannelSeeting}. It can be noted that the results verified on this channel dataset have strong guiding significance for further research on the future communication system design and protocol specification.

%\begin{table}[!htbp]\footnotesize
%\centering
%\caption{Channel and dataset parameter settings}
%\label{tabChannelSeeting}
%\begin{tabular}{|c|c|c|}
%\hline
% Parameter & Symbol & Value \\
%\hline
%Channel model & \diagbox{}{}  &  UMa \& NLos\\
% \hline
%Carriers frequency &$F_{\rm c}$&  3.5GHz\\
% \hline
%Subcarrier spacing & $B_{\rm sc}$  &  15KHz\\ 
% \hline
%Number of RB &  $N_{\rm RB}$  &  48\\
%\hline
% Number of tx antennas & $N_{\rm t}$ & 4 \\
% \hline
% Number of rx antennas  &  $N_{\rm r}$ & 4\\
% \hline
% Number of clusters & $N_{\rm d}$ & 24\\
% \hline
% Number of UEs in training set & $N_{\rm train}$  &  3000 \\
% \hline
% Number of UEs in testing set & $N_{\rm test}$  & 400 \\
% \hline
% Number of sampling slots & $N_{\rm slot}$  &  200 \\
% \hline
% Number of interval slots & $T$  &  100\\
% \hline
%\end{tabular}
%\end{table}

\begin{table}[tbp]\small
\centering
\caption{Channel and dataset parameter settings}
\label{tabChannelSeeting}
\begin{tabular}{|c|c|}
\hline
 Parameter  & Value \\
\hline
Channel model  &  UMa \& NLoS\\
 \hline
Carrier frequency $F_{\rm c}$ &  3.5GHz\\
 \hline
Subcarrier spacing  $B_{\rm sc}$  &  15KHz\\ 
 \hline
Number of resource block $N_{\rm RB}$  &  48\\
\hline
 Number of Tx antennas $N_{\rm t}$ & 4 \\
 \hline
 Number of Rx antennas $N_{\rm r}$ & 4\\
 \hline
 Number of clusters $N_{\rm d}$ & 24\\
 \hline
 Number of cells $N_{\rm c}$ & 57\\
 \hline
 Number of UEs in training set $N_{\rm train}$  &  3000 \\
 \hline
 Number of UEs in testing set $N_{\rm test}$  & 400 \\
 \hline
 Number of sampling slots $N_{\rm slot}$  &  200 \\
 \hline
 Number of interval slots $T$  &  100\\
 \hline
\end{tabular}
\end{table}

\raggedbottom
\vspace{-10pt}
\section{Discussions on the Potential Enhancements on F-CSI Feedback Design}
\label{Enhancing Schemes}
Compared with images and sequences in traditional computer vision (CV) and natural language processing (NLP) problems, respectively, the channel data obtained from wireless communication has some unique characteristics modeled by multi-antenna and multipath, etc. These characteristics lead to novel and suitable methods to process the channel data and design the neural network. Indeed, beyond existing DL-based methods for F-CSI feedback, a series of interesting enhancing schemes for DL-based F-CSI feedback have appeared during the WAIC. In this section, the schemes are divided into three categories, i.e., i) channel data analysis and preprocessing, ii) neural network design and iii) quantization enhancement, and discussed in more detail.

\subsection{Channel Data Analysis and Preprocessing}
\label{Channel Data Analysis and Preprocessing}

\begin{figure}[tb]
\centering
\setlength{\abovecaptionskip}{-0.5cm}
\includegraphics[scale=0.68]{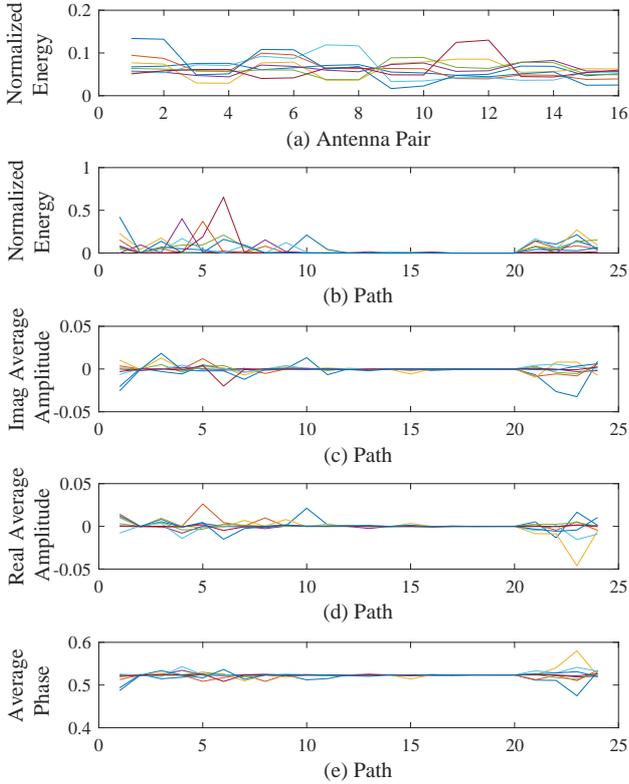}
\caption{Illustration of signal analysis in terms of antenna pair dimension and path dimension}
\label{dataAnalysis}
\end{figure}

\subsubsection{Data Analysis}
\label{Data Analysis}
Due to the unique characteristics of wireless channel data in comparison with image and sequence, the analysis of data is significant before model design. Fig. \ref{dataAnalysis} provides a detail signal analysis in terms of antenna and path dimensions, where $8$ samples of channel data are randomly chosen and depicted. Note that one antenna pair denotes a pair consisting of a transmitting antenna and a receiving antenna, and one path denotes a cluster consisting of multiple sub-paths as shown in Eq. (\ref{eqHsys}). Comparing with the energy distribution in Fig. \ref{dataAnalysis}(a) and Fig. \ref{dataAnalysis}(b), the path dimension possesses more obvious and regular energy distribution than antenna dimension, which indicates that several paths account for most of the energy. Furthermore, Fig. \ref{dataAnalysis}(c) to Fig. \ref{dataAnalysis}(e) show the amplitude of real and imaginary parts as well as the phase for each path. All these observations enlighten the scheme design introduced in the following part.

\subsubsection{Path Cutting}
\label{Path Cutting}
\begin{figure}[tb]
\centering
\includegraphics[scale=0.55]{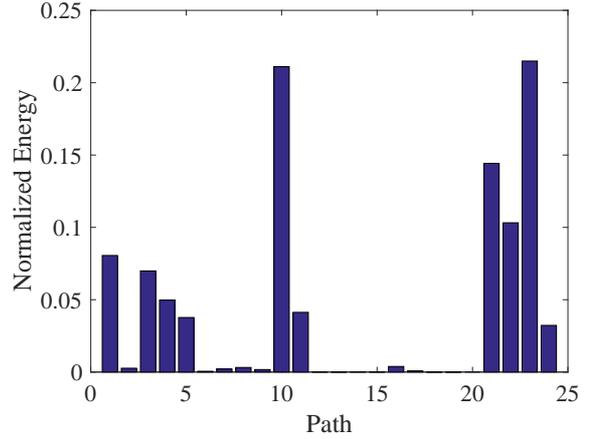}
\caption{Normalized energy for different paths.}
\label{pathCutting}
\end{figure}

Fig. \ref{pathCutting} shows the normalized energy for different paths of a sample in the channel data. It can be noticed that part of all paths accounts for most of the energy, and the energy of remaining paths is negligible in comparison. As the paths with higher energy carry the main information of the channel, path cutting, i.e., sorting the paths by energy, discarding the paths with lower energy and reserving paths with higher energy for compressing, can greatly reduce the size of the channel image to be compressed. The retaining paths with main information is enough for channel reconstruction at receiver. Taking Fig. \ref{pathCutting} as an example, $10$ paths acconting for main energy can be retained for compressing with $14$ paths with low energy discarded. Moreover, since the indices of the paths acconting for main energy in different sample is different, it need to take additional overhead to compress and feedback the indices so that the decompression process at the receiver can be executed correctly. 
%However, it is negligible in comparison with the overhead caused by compressing the discarded paths.

\subsubsection{Domain Transforming}
\label{Domain Transforming}
As data analysis in \ref{Data Analysis}, the energy distribution in antenna pair dimension is chaotic and lack of explicit features. However, the antenna-delay domain channel matrix can be transformed into angular-delay domain using Discrete Fourier Transform (DFT), in which the channel matrix exhibits a sparse feature \cite{wen2018deep}. That is, the elements of channel matrix contain only a small fraction of large components and the other components are close to zero. In addition to the transformation from the antenna-delay domain to the angular-delay domain, the channel matrix can also be transformed from antenna-delay domain to antenna-frequency domain. This transformation make the channel data more obvious explicit characteristics.

\begin{figure}[tb]
\centering
\includegraphics[scale=0.4]{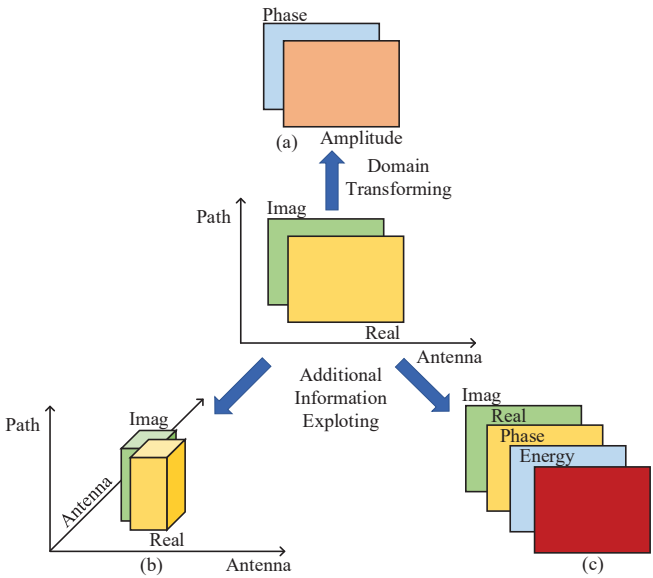}
\vspace{6pt}
\caption{Domain transforming and additional information exploting.}
\label{domainTransforming}
\end{figure}

In terms of processing the plural signal, existing DL network for F-CSI feedback often regard the real and imaginary parts of channel matrix as two feature maps \cite{vankayala2020optimizing}. Optionally, as shown in Fig. \ref{domainTransforming} (a), one can also transform the complex domain into the amplitude-phase domain treating the amplitude and phase of wireless channel matrix as two feature maps.

\vspace{-15pt}
\subsubsection{Additional Information Exploting}
\label{Additional Information Exploting}
The existing DL networks for F-CSI feedback \cite{wen2018deep,sun2020ancinet, lu2020multi,  cao2021lightweight, wang2018deep} shape the input data as the real and imaginary parts and then rely on the neural networks to extract implicit feature. However, according to the data analysis in section \ref{Data Analysis}, more features in channel data, such as energy, amplitude and phase, can be extracted manually before being input to the network, which can be exploited as explicit features for network to extract. As shown in Fig. \ref{domainTransforming} (c), one can optionally convert the input sample from two channels to four channels, where the phase and energy channels are the additional explicit feature information.

Moreover, shape of the input data can be guided by the pattern of the antenna pair. For example, when the antenna pair possesses the pattern with $1\times16$, reshaping the input channel matrix as $24$(paths)$\times16$(antenna pairs)$\times2$(real and imaginary channels) is appropriate for 2D convolution layer to process. While when the antenna pair possesses the pattern with $4\times4$, one can optionally reshape the input channel matrix as $24$(paths)$\times4$(antennas)$\times4$(antennas)$\times2$(real and imaginary channels) (as shown in Fig. \ref{domainTransforming} (b)) and exploit 3D convolution layer to extract the features of the antenna pair dimension.

\subsubsection{Data Augmentation}
\label{Data Augmentation}
Space transformation and color transformation are optional methods for data augmentation in CV. Since the channel data does not have the color feature similar to the image, it seems to be difficult for color transformation being directly applied to channel data and needs more study and discussion. However, the flipping and translating among the antenna pair dimension are potential methods as the channel data has certain spatial feature in the antenna pair dimension. Moreover, Cutout \cite{devries2017improved}, Mixup \cite{zhang2017mixup} and CutMix \cite{yun2019cutmix} can also optionally be utilized for data augmentation and alleviate the problem of over-fitting.

\subsection{Neural Network Design}
\label{Network Design}
As the core of DL, the type of data often determines the design of network model. For example, the strong ability of CNN on local feature extraction is suitable for image data processing in CV. As comparison, RNN are preferred for sequential data in NLP. Different from CV and NLP, the form and feature of data generated in wireless communication has a certain degree of particularity such as the plural signal and physical spatial features modeled by wireless transmission environment. Therefore, specific neural network model should be designed carefully for wireless communication, which includes two kinds of directions. Firstly, considering the maturity of DL in CV and NLP, several wireless communication problems can be mapped to these two fields and then existing and successful network models can be exploited. Secondly, according to the form and feature of wireless communication data, the specific modification on network design requires further exploration. In this section, three kinds of network design are listed for F-CSI feedback, including fully connected network (FCN), CNN and Transformer based models. Note that the examples introduced in this section are modified or design in the 1st WAIC.

\begin{figure}[tb]
\centering
\includegraphics[scale=0.4]{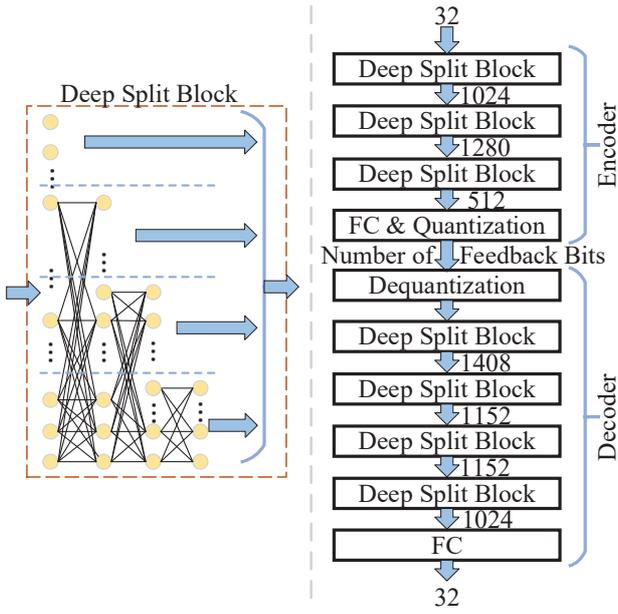}
\vspace{6pt}
\caption{Deep Split \cite{sarawgi2020have} FCN for F-CSI feedback.}
\vspace{12pt}
\label{deepSplit}
\end{figure}

\subsubsection{Fully Connected Neural Network Based Model}
\label{Fully Connected Network Based Model}
One of the optional designs considers the utilization of the FCN to solve the problem of F-CSI compression and recovery. Since the wireless environment becomes more and more complex with the scale of antennas is getting larger and larger, the channel data matrix exhibits certain nonlinear characteristics. Therefore, 
%Similar to the mathematical modeling method in current wireless communication systems, 
FCN regards the F-CSI compression and recovery as a nonlinear fitting problem, which is benefited from the multiple hidden layers in FCN.

For example, an FCN based model is illustrated in  Fig. \ref{deepSplit}, where the encoder and decoder both consist of a series of specific FCNs, named as Deep Split Blocks \cite{sarawgi2020have}. The model regards each path as the input with dimension $32$ ($2$ real and imaginary parts $\times 16$ antenna pairs) and multiple paths in one channel sample share the same model with fixed weights. The Deep Split Block is composed of $4$ fully connected (FC) layers, where parts of output nodes in each layer is directly connected to the final output. This triangular structure is able to make the network wide and deep at the same time with limited parameters. Specifically, the output dimensions of Deep Split Blocks can be set as \{$1024$, $1280$, $512$\} and \{$1408$, $1152$, $1152$ and $1024$\} for encoder and decoder, respcetively. Moreover, besides the Deep Split Block, one extra FC layer is used at the tail of encoder and decoder to adapt to the quantization bit and channel dimension, respectively. It is noted that in addition to the model that processes each path alone, multiple paths can also be treated together and jointly processed as the input of a neural network, which will be described int the following sections. 

\begin{figure}[tb]
\centering
\includegraphics[scale=0.45]{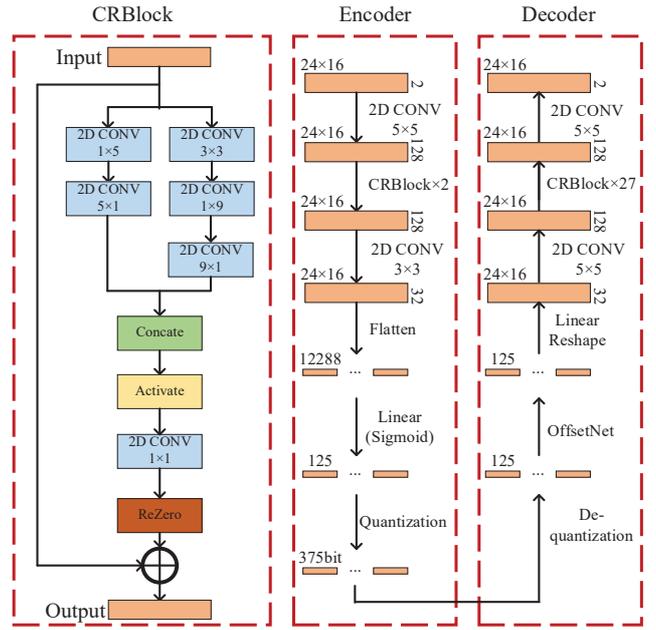}
\caption{CNN based model exploiting CRBlock modified CRNet in \cite{cao2021lightweight}.}
\label{CRNet}
\end{figure}

\subsubsection{Convolutional Neural Network Based Model}
\label{Convolutional Neural Network Based Model}
Mapping the F-CSI feedback to the image compression and recovery problem is considered in \cite{wen2018deep,sun2020ancinet, lu2020multi,  cao2021lightweight, wang2018deep}, where the channel matrix can be regarded as an image and processed by widely utilized CNNs in CV field. As one of the optional CNN based model, CRNet is proposed in \cite{cao2021lightweight}, where parallel network paths with different convolution kernel sizes are arranged to extract features on different scales, which can be concatenated in the following layers.

For example, a CRNet based model is given in Fig. \ref{CRNet}, wherein the encoder and decoder are both designed with CRBlock. The channel matrix is treaded as an input image with size $24$(paths)$\times16$(antenna pairs)$\times2$(real and imaginary channels). For the CRBlock, input channel image will pass through two parallel paths. One path providing large resolution view is made up of three serial convolution layers whose kernal sizes are $3\times3$, $1\times9$ and $9\times1$, respectively. While another path with much smaller resolution contains two serial convolution layers with kernal sizes $1\times5$ and $5\times1$, respectively. Finally, outputs from two paths are concatenated and merged by a $1\times1$ convolution layer. Moreover, a ReZero block \cite{bachlechner2020rezero} is also introduced in CRBlock, which modifies the residual block by adding one learnable eight $\alpha$ to the features obtaind from convolution, i.e., 
\begin{equation}
y_{\rm ReZero} = x_{\rm ReZero}+\alpha\mathcal{F}(x_{\rm ReZero})
\label{reZero}
\text{,}
\end{equation}
where $\mathcal{F}(\cdot)$ represents the convolutional layers, $x_{\rm ReZero}$ and $y_{\rm ReZero}$ denotes the input and output of ReZero model, respectively. The weight $\alpha$ is initialized as $0$ to reduce the difficulty and hence improve the speed of training. Finally, a residual structure is utilized, where the output is obtained by adding the input of CRBlock and the output of ReZero. Moreover, two convolution layers with $3\times3$ kernal, a FC layer and a quantization layer are added at the tail of encoder to scale down the feature with the limited compression ratio and feedback bits. Specifically, for the decoder, an OffsetNet \cite{chen2019novel} is introduced to reduce the quantization error, which will be illustrated in detail in following Section \ref{Learnable Quantization Error Offset}. 

Generally, features between adjacent paths and elements are extracted by CRBlock considering same weights on each feature channel. Meanwhile, in order to  further improve the feature extraction performance, another structure, referred to as SE-Net \cite{hu2018squeeze}, is utilized in WAIC, which implements the attention mechanism on different feature channels. Specifically, SE-Net performs squeeze firstly and obtain one vector by conducting global average pooling for each feature channel. Secondly, the vector is inputed to an FC layer to get the attention weight value for each feature channel. Finally, each feature channel is multiplied by its corresponding value to obtain a new feature map. In fact, most of CNN based structures can be enhanced with SE-Net for feature recalibration.

\subsubsection{Transformer Based Model}
\label{Transformer Based Model}
\begin{figure}[tb]
\centering
\includegraphics[scale=0.8]{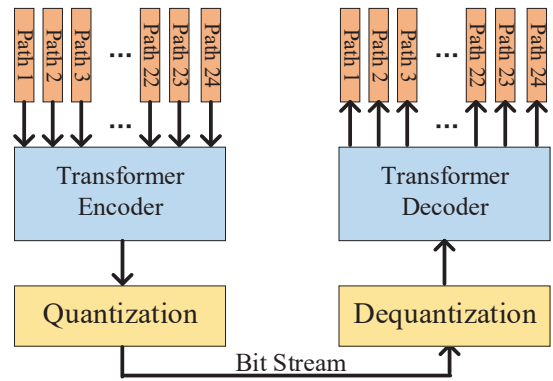}
\caption{Transformer implementation \cite{vaswani2017attention} for F-CSI feedback.}
\label{transformer}
\end{figure}

Besides CNN based models which treat channel as an image, RNN and Transformer based models are also proposed \cite{vaswani2017attention}, where the channels with multiple paths are regarded as an sequence input, which is similar to the sequential processing issue in NLP. For example, a Transformer based model is shown in Fig. \ref{transformer}, which provides three potentials and advantages for F-CSI feedback. Firstly, its natural encoder-decoder structure is quite suitable for implementing at the UE and BS respectively. Secondly, different paths or antennas are treated as the sequence inputs, which benefit the extraction of physical characteristics of wireless channel. Thirdly, the self-attention block is utilized in Transformer so that the correlation information between different paths and antennas can be further explored.

\subsection{Quantization enhancement}
\label{Quantization enhancement}
The feature vector compressed by an AI encoder needs to be fed back to receiver in form of a bit stream. As shown in Fig. \ref{fig2}, the process transforming feature vector to bitstream and bitstream to feature vector are generally completed by a quantizer. However, quantization generally brings loss of accuracy, and different quantization strategy choices will also affect the performance of F-CSI feedback. A good quantizer design can use a small number of feedback bits to accurately express the compressed channel data.
%Since the information in a wireless communication system is transmitted in the form of bit stream, the quantization process is indispensable. However, given a constant number of feedback overhead bits $L_{\rm overhead}$, the larger number of quantization bits is, the shorter feature vector fed back is and vice versa. Therefore there is a trade-off between the length of the feature vector $L_{\rm vector}$ and the number of quantization bits $B$, i.e., 
%\begin{equation}
%L_{\rm overhead} = B\times L_{\rm vector}
%\label{overheadEq}
%\text{.}
%\end{equation}
Therefore, the quantization module is critical to the performance of the entire F-CSI feedback framework. In addition to the traditional uniform scalar quantification scheme, there are a series of quantization enhancements can be implemented. In this section, we will introduce five representative enhancing schemes including i) scalar quantization, ii) vector quantization, iii) path-level non-uniform quantization, iv) learnable quantization error offset and v) derivable quantized step function.
\begin{algorithm}[t]
\caption{\large{Alternating iterative optimization for quantization interval and quantization level}}
\label{algLevelSearch}
\begin{algorithmic}
\STATE \hspace*{-5mm} \textbf{Input}: \\Channel matrix $\mathbf{H}$, network of encoder and decoder $\mathcal{E}$ and $\mathcal{D}$, quantization interval $\mathbf{x}_0$ and quantization level $\hat{\mathbf{x}}_0$.
\STATE \hspace*{-5mm} \textbf{Output}: \\Quantization interval $\mathbf{x}$ and quantization level $\hat{\mathbf{x}}$ of non-uniform quantizer.
\STATE \hspace*{-5mm} \textbf{Initialization}: \\Freeze the pre-trained weights of $\mathcal{E}$ and $\mathcal{D}$ and initialize the quantization interval $\mathbf{x}_0$ and quantization level $\hat{\mathbf{x}}_0$ as the parameters of uniform quantizer.
\STATE \hspace*{-5mm} \textbf{Optimization goal}: \\Minimize NMSE.
\STATE \hspace*{-5mm} \textbf{Step 1: } Freeze the quantization level $\hat{\mathbf{x}}$ and search the quantization interval $\mathbf{x}$ until the NMSE is locally optimal;
\STATE \hspace*{-5mm} \textbf{Step 2: } Freeze the quantization interval $\mathbf{x}$  and search the quantization level $\hat{\mathbf{x}}$ until the NMSE is locally optimal;
\STATE \hspace*{-5mm} \textbf{Step 3:} Iterate Steps 1 and 2 until the NMSE converges.
\end{algorithmic}
\end{algorithm}

\subsubsection{Scalar Quantization}
\label{Scalar Quantization}

\emph{a) Non-adaptive Scalar Quantization:} \\
Non-adaptive scalar quantization quantizes each element in feature vector independently. The amplitude of the element in a feature vector is divided into several quantization intervals, and the amplitude value falling in the quantization interval is quantized and output as a corresponding quantization level. The uniform and non-uniform quantification consider the uniform and non-uniform quantization interval, respectively. Since the distribution of amplitude of element in the feature vector is not uniform, the non-uniform quantization such as $\mu$-law and A-law can be applied, which can cope with non-uniform distribution to a certain extent.\\
\emph{b) Adaptive Scalar Quantization:} \\
Because of the unanalyzability of the amplitude distribution of feature vector output by the feature extraction network at the encoder, it is difficult to manually design a suitable and fixed non-uniform quantization interval. As shown in algorithm \ref{algLevelSearch}, one potential scheme used in the competition, namely adaptive scalar quantization, optimizes the quantization interval and quantization level utilizitng the prior information of quantitative distribution of data. The algorithm train the quantizer separately by freezing the other parameters of the network and iteratively optimizing quantization interval and quantization level.

\vspace{-5pt}
\subsubsection{Vector Quantization}
\label{Vector Quantization}
\begin{figure}[tb]
\centering
\includegraphics[scale=0.75]{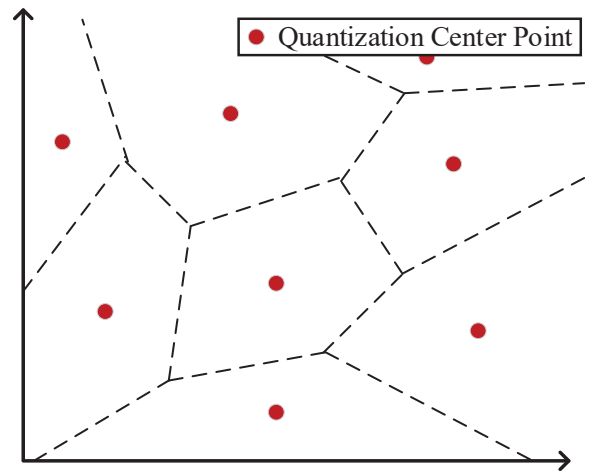}
\caption{Vector quantization space taking two-dimensional space as an example.}
\label{vectorQuan}
\end{figure}

Compared with scalar quantization for quantizing each element in a feature vector separately, vector quantization divides the feature vector into multiple sub-vectors for quantization, regarding some points in the vector space spanned by the sub-vectors of the feature vector as the quantifing center (as shown in Fig. \ref{vectorQuan}). Morevoer, the search of quantization interval and the quantization center vector quantization is regarded as a clustering problem, which can be solved by DL and make good use of the prior distribution information of the data. Similar to the adaptive scalar quantization scheme in section \ref{Scalar Quantization} b), the quantization center can be obtained from training phase and be fixed during inference phase as the shared knowledge between BS and UE. 
%Another more radical design considers updating the quantization center for each feedback using online learning. This method can obtain quantizer parameters more suitable for a single sample in the encoding stage. However, due to the information asymmetry between the UE and the BS for the constantly changing quantization center, the feedback of the quantization center information is inevitable and leads to the additional overhead, i.e.,
%\begin{equation}
%L_{\rm overhead} = L_{\rm info} + L_{\rm quan}
%\label{overheadEq2}
%\text{,}
%\end{equation}
%where $L_{\rm overhead}$, $L_{\rm info}$ and $L_{\rm quan}$ denote the length of overhead in total, length of overhead for feeding back channel feature and length of overhead for feeding back quantization parameter, respectively. Note that the more quantization center points, the better the quantization performance but the larger the overhead $L_{\rm quan}$. Therefore, given a fixed feedback overhead for quantization parameter $L_{\rm quan}$, there is a trade-off between the number of quantization center points and quantization performance.

\subsubsection{Path-level Non-uniform Quantization}
\label{Path-level Non-uniform Quantization}
According to the data analysis in section \ref{Data Analysis}, the amplitude and energy of differetn path in the channel data is quite different. A path-level quantization could be considered. From the perspective of amplitude distribution, the adaptability to each path is relatively low if the same quantization parameters including quantization interval, quantization level, and quantization bit number are used for different paths. Meanwhile, it can be noticed that paths with larger energy often carry more feature information of channel images and vice versa. Applying the same number of quantization bits for different paths will cause large quantization errors (for the paths with larger power) and low coding efficiency (for paths with lower energy). In this case, path-level non-uniform quantization can be applied by allocating different number of quantization bits for each path. Optionally, the number of quantization bits of different paths can be manually set accroding to data analysis or learned by neural network in the training phase. 
%Another method can also consider the utilization of online learning in the inference phase to update the number of quantized bits in real time. However, similar to the scheme utilizing online learning in section \ref{Vector Quantization}, it still needs additional overhead to feedback the information of quantization bits number.

\subsubsection{Learnable Quantization Error Offset}
\label{Learnable Quantization Error Offset}
\begin{figure}[tb]
\centering
\includegraphics[scale=0.28]{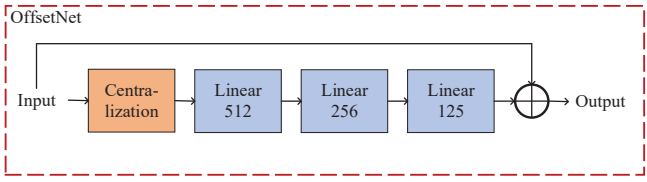}
\caption{Illustration of structure of OffsetNet.}
\label{offsetNet}
\end{figure}

Quantization error is inevitable and has a negative impact on performance of F-CSI feedback. As a potential solution, OffsetNet (as shown in Fig. \ref{offsetNet}) \cite{chen2019novel} can be implemented after dequantization at decoder to offset the quantization errors using neural networks. The feature vector with and without quantization error can be the input and output of the OffsetNet for training, respectively. OffsetNet has a simple residual structure including centralization module and three FC layers. The vector is first passed through the centralization module to subtract the mean value, in order to improve the convergence speed of following FC layers. Next, the following FC layers conduct the error feature learning, and the error offset are completed by residual operation at last.

\subsubsection{Derivable Quantized Step Function}
\label{Derivable Quantized Step Function}
\begin{figure}[bt]
\centering
\includegraphics[scale=0.38]{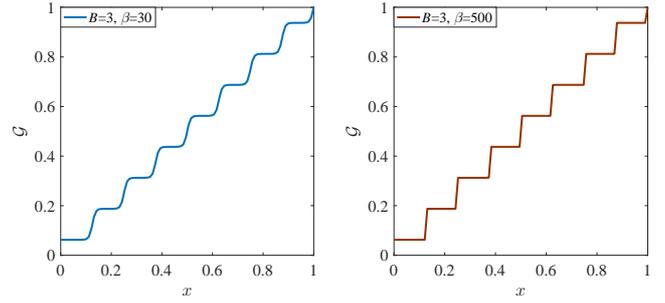}
\caption{Approximate step function $\mathcal{G}$ with the number of quantization bits $B=3$ and the approximation coefficient $\beta=\{30,500\}$.}
\label{stepFunction}
\end{figure}

Most of the existing networks for F-CSI feedback include non-derivable quantization and dequantization layers, through which the gradient cannot be back propagated. To solve this problem, one idea is bypassing the quantization and dequantization layer when propagating the gradient in training phase. However, this solution will cause gradient propagation error and reduce the training effect. In order to solve the problem of gradient propagation error, a potential solution utilizes the sum of several sigmoid functions to form a approximative step function which can replace uniform quantization and dequantization layers in training phase, i.e.,
\begin{equation}
\begin{split}
&\mathcal{G}(x,B,\beta)= \\
&\frac{0.5+\sum\limits_{i = -2^{B-1}}^{2^{B-1}-1}\mathrm{sigmoid}\left(\beta(2^{B}x-i-1-2^{B-1})\right)}{2^{B}}
\end{split}
\label{stepFunctionEq}
\text{,}
\end{equation}
where $x$, $B$ and $\beta$ denote the input data, number of quantization bits and approximation coefficients, respectively. $\mathrm{sigmoid}(\cdot)$ represents the sigmoid function. Fig. \ref{stepFunction} shows the function curve with the number of quantization bits $B=3$ for varying the approximation coefficient $\beta$. It can be noticed that the curve with $\beta=30$ is smoother than the curve with $\beta=500$. The approximate function makes it possible that the gradient propagating across the non-directed quantization and dequantization layer during the back propagation process, leading to a more accurate back propagation of the gradient. Note that approximate step function only equivalently replaces the quantization and dequantization layers in training phase to improve the accuracy of gradient transfer. 

\begin{table*}[bt]
\Large
\caption{DL-based F-CSI feedback solutions with different schemes.}
\begin{threeparttable}
\resizebox{\textwidth}{!}{
%\begin{tabular}{|Y|Y|Y|Y|Y|Y|Y|Y|Y|Y|Y|Y|Y|Y|}
\renewcommand{\arraystretch}{1.06}
\begin{tabular}{|m{0.8cm}<{\centering}|m{0.8cm}<{\centering}|m{1.9cm}<{\centering}|m{1.9cm}<{\centering}|m{1.9cm}<{\centering}|m{1.9cm}<{\centering}|m{1.9cm}<{\centering}|m{1.9cm}<{\centering}|m{1.9cm}<{\centering}|m{1.9cm}<{\centering}|m{1.9cm}<{\centering}|m{1.9cm}<{\centering}|m{1.9cm}<{\centering}|m{1.9cm}<{\centering}|}
%\begin{tabular}{|m{1.2cm}<{\centering}|m{1.9cm}<{\centering}|m{1.9cm}<{\centering}|m{1.9cm}<{\centering}|m{1.9cm}<{\centering}|m{1.9cm}<{\centering}|m{1.9cm}<{\centering}|m{1.9cm}<{\centering}|m{1.9cm}<{\centering}|m{1.9cm}<{\centering}|m{1.9cm}<{\centering}|m{1.9cm}<{\centering}|m{1.9cm}<{\centering}|m{1.9cm}<{\centering}|}
\hline
\multirow{3}{*}{\begin{tabular}[m]{@{}c@{}}\\Solu-\\ tion\\ ID\end{tabular}}&\multirow{3}{*}{\begin{tabular}[c]{@{}c@{}}\\ \\ Bits\tnote{1}\end{tabular}}&\multicolumn{12}{c|}{ Scheme}                                                                                                                                                                                                                                                                                                                                                \\
\cline{3-14}
&& \multicolumn{4}{c|}{Data Preprocessing}                                                                                                                                                                                                                             & \multicolumn{3}{c|}{Backbone}   & \multicolumn{5}{c|}{Quantization Enhancement}                                                                                                                                                                                                                                                                                                                                                 \\ \cline{3-14}
   & & \begin{tabular}[c]{@{}c@{}} Path\\  Cutting\end{tabular} & \begin{tabular}[c]{@{}c@{}}Domain\\Trans-\\forming\end{tabular} & \begin{tabular}[c]{@{}c@{}}Additional\\ Info.\\ Exploiting\end{tabular} & \begin{tabular}[c]{@{}c@{}}Data\\ Aug-\\mentation\end{tabular} & \begin{tabular}[c]{@{}c@{}}\\ FCN \end{tabular}    & \begin{tabular}[c]{@{}c@{}}\\ CNN \end{tabular}        &\begin{tabular}[c]{@{}c@{}}Trans-\\former\end{tabular} & \begin{tabular}[c]{@{}c@{}}Adaptive\\ Scalar\\ Quan.\end{tabular} & \begin{tabular}[c]{@{}c@{}}Vector\\ Quan.\end{tabular} & \begin{tabular}[c]{@{}c@{}}Path-level\\ Non-\\uniform\\ Quan.\end{tabular} & \begin{tabular}[c]{@{}c@{}}Learnable\\ Quan.\\ Error \\Offset\end{tabular} & \begin{tabular}[c]{@{}c@{}}Derivable\\ Quantized\\ Step \\Function\end{tabular} \\ \hline
1& 286                                  &  $\surd$                                                &                                                               &                                                                              &                                                             &         &         & $\surd$      &                                                                           &  $\surd$                                                       &  $\surd$                                                                         &                                                                                 &                                                                          \\ [0.3cm] \hline
2&  354                                  &  $\surd$                                                &                                                               &                                                                              &                                                             & $\surd$ &         &             &                                                                           & $\surd$                                                       & $\surd$                                                                         &                                                                                 &                                                                               \\[0.3cm] \hline
 3& 363                                   & $\surd$                                                & $\surd$                                                       &                                                                              &                                                             &         & $\surd$ &             &                                                                           &                                                               &                                                                                 &                                                                                 &                                                                               \\[0.3cm] \hline
4&369                                   &                                                        &                                                               & $\surd$                                                                      &                                                             &         & $\surd$ &             &                                                                   &                                                               &                                                                                 &                                                                                 &                                                                               \\[0.3cm] \hline
 5&372                                   &                                                        &                                                               &                                                                              &                                                             &         & $\surd$ &             & $\surd$                                                                   &                                                               &                                                                                 &                                                                                 &                                                                               \\[0.3cm] \hline
6& 375                                   &                                                        &                                                               &                                                                              & $\surd$                                                     &         & $\surd$ &             &                                                                           &                                                               &                                                                                 &$\surd$                                                                         &                                                                               \\ [0.3cm]\hline
7& 378                                   &                                                        &                                                               &                                                                              &$\surd$                                                     &         & $\surd$ &             &                                                                           &                                                               &                                                                                 & $\surd$                                                                         &                                                                               \\[0.3cm] \hline
 8& 381                                  &                                                        &                                                               &                                                                              &                                                             &         & $\surd$ &             &                                                                           &                                                               &                                                                                 &                                                                                 &                                                                               \\ [0.3cm]\hline
 9& 384                                 &                                                        &                                                               &                                                                              &                                                             &         &         & $\surd$     &                                                                           &                                                               &                                                                                 &                                                                                 &                                                                               \\[0.3cm] \hline
 10& 384                                 &                                                        &                                                               &                                                                              &                                                             &         &$\surd$ &             &                                                                  &                                                               &                                                                                 &                                                                                 & $\surd$                                                                       \\[0.3cm] \hline

\end{tabular}
}
\begin{tablenotes}
    \normalsize \item[1] Note that the number of feedback bits satisfies that $\rm{NMSE} = \rm{E}\left\{||\mathbf{H}'-\mathbf{H}||_2^2/||\mathbf{H}||_2^2\right\}\leq0.1$.
\end{tablenotes}
\end{threeparttable}
\label{Capability composition of participating models table}
\end{table*}

\section{F-CSI Feedback Solutions and Challanges}
\label{Capability Composition of Participating Models}
There are more than $900$ teams registered for the 1st WAIC. Among their work, solutions from top $10$ teams with various backbone choices are listed in Table \ref{Capability composition of participating models table}. The original channel matrix with a size of $24$ (paths) $\times$ $16$ (antenna pairs) $\times2$ (real and imaginary parts) $\times$ $32$ ($32$-bit floating point) $= 24576$ bits can be compressed to $286$ bits with NMSE $\leq 0.1$ utilizing the best solution. It can be noticed that with proper training method, both FCN, CNN and Transformer based models can achieve good performance from the perspective of F-CSI feedback overhead. Secondly, data preprocessing plays an important role for overhead reduction. By removing unnecessary data with path cutting in model training phase, higher F-CSI compression ratio can be obtained. Other approaches, such as domain transforming, additional information exploiting and data augmentation, are also adopted in different solutions. Thirdly, proper quantization strategy is quite crucial for F-CSI compression. Vector quantization and path-level non-uniform quantization can significantly improve the quantization performance with lower feedback bits. In addition to the enhancement methods listed in Table \ref{Capability composition of participating models table}, some useful training tricks, such as parameter tuning, warm-up and pre-training can be used to improve the performance of given networks, which is not included in this paper and can be further studied in the future.

%
%
%\vspace{-0.25cm}
%\section{Challenge}
%\label{Challenge}
Even though DL-based solutions can achieve excellent performance for F-CSI feedback problem with complex 3GPP channel, it is still the first step for exploring AI model-based solution for wireless communication system. Whether AI based solutions could be widely used in commercial wireless communication systems is still required further study. Firstly, how to define and construct the datasets, how to train and refine the AI model, how to handle and utilize the over-fitting and generalization features of AI solutions in wireless communication systems need more research. Meanwhile, proper link-level and system-level simulations should be considered to verify the performance gain of AI solutions compared to traditional solutions under given conditions as well as limited cost. In addition, in order to widely deploy AI solutions in commercial wireless communication systems, more issues on specification impacts should be studied considering different AI solutions on various use cases.

\section{CONCLUSION}
\label{CONCLUSION}

In this paper, we first give a description to the framework of F-CSI feedback and its corresponding channel environment involved in the 1st WAIC. Then the enhancing schemes for DL-based F-CSI feedback including i) channel data analysis and preprocessing, ii) neural network design and iii) quantization enhancement are introduced. After that, the solutions composed of different enhancing schemes are listed. Finally, we analyze and prospect the challenge on the design of AI-based wireless communication systems. As a representative research on the application of DL in wireless communication systems, DL-based F-CSI feedback is a good beginning for AI based wireless communication. Here we sincerely hope that this article can broaden the thinking and insights for interested researchers.

\section*{ACKNOWLEDGEMENT}
\label{ACKNOWLEDGEMENT}
We thank China Academy of Information and Communications Technology,  Guangdong OPPO Mobile Telecommunications Corp., Ltd, and National Mobile Communications Research Laboratory of Southeast University for their help and support on the research and the holding of the 1st WAIC. We also would like to express our gratitude to all contestants for their participation and sharing.

\bibliographystyle{gbt7714-numerical}
\bibliography{main}

\begin{thebibliography}{28}
\providecommand{\natexlab}[1]{#1}
\providecommand{\url}[1]{#1}
\expandafter\ifx\csname urlstyle\endcsname\relax\else
  \urlstyle{same}\fi
\expandafter\ifx\csname href\endcsname\relax
  \DeclareUrlCommand\doi{\urlstyle{rm}}\else
  \providecommand\doi[1]{\href{https://doi.org/#1}{\nolinkurl{#1}}}\fi

\bibitem[3GPP(2020{\natexlab{a}})]{1}
3GPP.
\newblock {3GPP TS} 38.212 v16.1.0, 3rd generation partnership project;
  technical specification group radio access network; {NR}; multiplexing and
  channel coding (release 16)\allowbreak[J].
\newblock Tech. Rep., 2020.

\bibitem[3GPP(2020{\natexlab{b}})]{2}
3GPP.
\newblock {3GPP TS} 38.214 v16.1.0 3rd generation partnership project;
  technical specification group radio access network; {NR}; physical layer
  procedures for data (release 16)\allowbreak[J].
\newblock Tech. Rep., 2020.

\bibitem[3GPP(2020{\natexlab{c}})]{3}
3GPP.
\newblock {3GPP TS} 38.331 v16.0.0 3rd generation partnership project;
  technical specification group radio access network; {NR}; radio resource
  control ({RRC}) protocol specification (release 16)\allowbreak[J].
\newblock Tech. Rep., 2020.

\bibitem[Wang et~al.(2017)Wang, Wen, Wang, Gao, Jiang, and Jin]{wang2017deep}
WANG~T, WEN~C~K, WANG~H, et~al.
\newblock Deep learning for wireless physical layer: opportunities and
  challenges\allowbreak[J].
\newblock China Communications, 2017, 14\penalty0 (11):\penalty0 92-111.

\bibitem[Wen et~al.(2018)Wen, Shih, and Jin]{wen2018deep}
WEN~C~K, SHIH~W~T, JIN~S.
\newblock Deep learning for massive {MIMO} {CSI} feedback\allowbreak[J].
\newblock IEEE Wireless Communications Letters, 2018, 7\penalty0 (5):\penalty0
  748-751.

\bibitem[Sun et~al.(2020)Sun, Xu, Fan, Li, and Karagiannidis]{sun2020ancinet}
SUN~Y, XU~W, FAN~L, et~al.
\newblock Ancinet: An efficient deep learning approach for feedback compression
  of estimated {CSI} in massive {MIMO} systems\allowbreak[J].
\newblock IEEE Wireless Communications Letters, 2020, 9\penalty0 (12):\penalty0
  2192-2196.

\bibitem[Lu et~al.(2020)Lu, Wang, and Song]{lu2020multi}
LU~Z, WANG~J, SONG~J.
\newblock Multi-resolution {CSI} feedback with deep learning in massive {MIMO}
  system\allowbreak[C]//\allowbreak
ICC 2020-2020 IEEE International Conference on Communications (ICC).
\newblock [S.l.]: IEEE, 2020: 1-6.

\bibitem[Chen et~al.(2020)Chen, Guo, Wen, Jin, Li, Wang, and Hou]{chen2020deep}
CHEN~T, GUO~J, WEN~C~K, et~al.
\newblock Deep learning for joint channel estimation and feedback in massive
  {MIMO} systems\allowbreak[J].
\newblock arXiv preprint arXiv:2011.07242, 2020.

\bibitem[Mashhadi et~al.(2020)Mashhadi, Yang, and
  G{\"u}nd{\"u}z]{mashhadi2020distributed}
MASHHADI~M~B, YANG~Q, G{\"U}ND{\"U}Z~D.
\newblock Distributed deep convolutional compression for massive {MIMO} {CSI}
  feedback\allowbreak[J].
\newblock IEEE Transactions on Wireless Communications, 2020.

\bibitem[Cao et~al.(2021)Cao, Shih, Guo, Wen, and Jin]{cao2021lightweight}
CAO~Z, SHIH~W~T, GUO~J, et~al.
\newblock Lightweight convolutional neural networks for {CSI} feedback in
  massive {MIMO}\allowbreak[J].
\newblock IEEE Communications Letters, 2021.

\bibitem[Guo et~al.(2020{\natexlab{a}})Guo, Wen, and Jin]{guo2020deep}
GUO~J, WEN~C~K, JIN~S.
\newblock Deep learning-based {CSI} feedback for beamforming in single-and
  multi-cell massive {MIMO} systems\allowbreak[J].
\newblock IEEE Journal on Selected Areas in Communications, 2020.

\bibitem[Guo et~al.(2021)Guo, Wen, and Jin]{guo2021canet}
GUO~J, WEN~C~K, JIN~S.
\newblock Canet: Uplink-aided downlink channel acquisition in {FDD} massive
  {MIMO} using deep learning\allowbreak[J].
\newblock arXiv preprint arXiv:2101.04377, 2021.

\bibitem[Lu et~al.(2018)Lu, Xu, Shen, Zhu, and Wang]{lu2018mimo}
LU~C, XU~W, SHEN~H, et~al.
\newblock {MIMO} channel information feedback using deep recurrent
  network\allowbreak[J].
\newblock IEEE Communications Letters, 2018, 23\penalty0 (1):\penalty0 188-191.

\bibitem[Wang et~al.(2018)Wang, Wen, Jin, and Li]{wang2018deep}
WANG~T, WEN~C~K, JIN~S, et~al.
\newblock Deep learning-based {CSI} feedback approach for time-varying massive
  {MIMO} channels\allowbreak[J].
\newblock IEEE Wireless Communications Letters, 2018, 8\penalty0 (2):\penalty0
  416-419.

\bibitem[Guo et~al.(2020{\natexlab{b}})Guo, Wen, Jin, and
  Li]{guo2020convolutional}
GUO~J, WEN~C~K, JIN~S, et~al.
\newblock Convolutional neural network-based multiple-rate compressive sensing
  for massive {MIMO} {CSI} feedback: design, simulation, and
  analysis\allowbreak[J].
\newblock IEEE Transactions on Wireless Communications, 2020, 19\penalty0
  (4):\penalty0 2827-2840.

\bibitem[Li et~al.(2020)Li and Wu]{li2020spatio}
LI~X, WU~H.
\newblock Spatio-temporal representation with deep neural recurrent network in
  {MIMO} {CSI} feedback\allowbreak[J].
\newblock IEEE Wireless Communications Letters, 2020, 9\penalty0 (5):\penalty0
  653-657.

\bibitem[Chen et~al.(2019)Chen, Guo, Jin, Wen, and Li]{chen2019novel}
CHEN~T, GUO~J, JIN~S, et~al.
\newblock A novel quantization method for deep learning-based massive {MIMO}
  {CSI} feedback\allowbreak[C]//\allowbreak
2019 IEEE Global Conference on Signal and Information Processing (GlobalSIP).
\newblock [S.l.]: IEEE, 2019: 1-5.

\bibitem[Lu et~al.(2019)Lu, Xu, Jin, and Wang]{lu2019bit}
LU~C, XU~W, JIN~S, et~al.
\newblock Bit-level optimized neural network for multi-antenna channel
  quantization\allowbreak[J].
\newblock IEEE Wireless Communications Letters, 2019, 9\penalty0 (1):\penalty0
  87-90.

\bibitem[Jiang et~al.(2017)Jiang, Tao, Liu, Ren, Guo, and Zhao]{jiang2017end}
JIANG~F, TAO~W, LIU~S, et~al.
\newblock An end-to-end compression framework based on convolutional neural
  networks\allowbreak[J].
\newblock IEEE Transactions on Circuits and Systems for Video Technology, 2017,
  28\penalty0 (10):\penalty0 3007-3018.

\bibitem[3GPP(2020{\natexlab{d}})]{4}
3GPP.
\newblock {3GPP TR} 38.901 v16.1.0 3rd generation partnership project;
  technical specification group radio access network; study on channel model
  for frequencies from 0.5 to 100 {GH}z (release 16)\allowbreak[J].
\newblock Tech. Rep., 2020.

\bibitem[Vankayala et~al.(2020)Vankayala, Kumar, and
  Kommineni]{vankayala2020optimizing}
VANKAYALA~S~K, KUMAR~S, KOMMINENI~I.
\newblock Optimizing deep learning based channel estimation using channel
  response arrangement\allowbreak[C]//\allowbreak
2020 IEEE International Conference on Electronics, Computing and Communication
  Technologies (CONECCT).
\newblock [S.l.]: IEEE, 2020: 1-5.

\bibitem[DeVries et~al.(2017)DeVries and Taylor]{devries2017improved}
DEVRIES~T, TAYLOR~G~W.
\newblock Improved regularization of convolutional neural networks with
  cutout\allowbreak[J].
\newblock arXiv preprint arXiv:1708.04552, 2017.

\bibitem[Zhang et~al.(2017)Zhang, Cisse, Dauphin, and
  Lopez-Paz]{zhang2017mixup}
ZHANG~H, CISSE~M, DAUPHIN~Y~N, et~al.
\newblock mixup: Beyond empirical risk minimization\allowbreak[J].
\newblock arXiv preprint arXiv:1710.09412, 2017.

\bibitem[Yun et~al.(2019)Yun, Han, Oh, Chun, Choe, and Yoo]{yun2019cutmix}
YUN~S, HAN~D, OH~S~J, et~al.
\newblock Cutmix: Regularization strategy to train strong classifiers with
  localizable features\allowbreak[C]//\allowbreak
Proceedings of the IEEE/CVF International Conference on Computer Vision.
\newblock [S.l.: s.n.], 2019: 6023-6032.

\bibitem[Sarawgi et~al.(2020)Sarawgi, Zulfikar, Khincha, and
  Maes]{sarawgi2020have}
SARAWGI~U, ZULFIKAR~W, KHINCHA~R, et~al.
\newblock Why have a unified predictive uncertainty? disentangling it using
  deep split ensembles\allowbreak[J].
\newblock arXiv preprint arXiv:2009.12406, 2020.

\bibitem[Bachlechner et~al.(2020)Bachlechner, Majumder, Mao, Cottrell, and
  McAuley]{bachlechner2020rezero}
BACHLECHNER~T, MAJUMDER~B~P, MAO~H~H, et~al.
\newblock Rezero is all you need: Fast convergence at large
  depth\allowbreak[J].
\newblock arXiv preprint arXiv:2003.04887, 2020.

\bibitem[Hu et~al.(2018)Hu, Shen, and Sun]{hu2018squeeze}
HU~J, SHEN~L, SUN~G.
\newblock Squeeze-and-excitation networks\allowbreak[C]//\allowbreak
Proceedings of the IEEE conference on computer vision and pattern recognition.
\newblock [S.l.: s.n.], 2018: 7132-7141.

\bibitem[Vaswani et~al.(2017)Vaswani, Shazeer, Parmar, Uszkoreit, Jones, Gomez,
  Kaiser, and Polosukhin]{vaswani2017attention}
VASWANI~A, SHAZEER~N, PARMAR~N, et~al.
\newblock Attention is all you need\allowbreak[J].
\newblock arXiv preprint arXiv:1706.03762, 2017.

\end{thebibliography}

\end{document}